\documentclass[conference,a4paper]{IEEEtran}
\usepackage[english]{babel}
\usepackage{amsmath,amssymb,amsfonts,mathrsfs,dsfont}

\DeclareMathOperator{\Tr}{Tr}
\newcommand{\id}{\mathds{1}}
\newcommand{\ket}[1]{|#1\rangle}
\usepackage{mathtools}
\mathtoolsset{showonlyrefs=true}
\usepackage{tikz}
\usetikzlibrary{positioning}
\usepackage{subfig}
\usepackage{adjustbox}
\renewcommand{\thetable}{\arabic{table}}

\begin{document}

\title{Perfect signaling among three parties\\violating predefined causal order}

\author{
  \IEEEauthorblockN{\"Amin Baumeler and Stefan Wolf}
  \IEEEauthorblockA{Faculty of Informatics\\
    Universit\`a della Svizzera italiana\\
    6900 Lugano, Switzerland\\
    Email: \{baumea,wolfs\}@usi.ch} 
}

%% To balance the two columns, you should reduce the text-height of
%% the last page using the following command:
%%%%%%%%%%%%%%%%%%%%%%%%%%%%%%%%%%%%%%%%%%%%%%%%%%%%%%%%%%%%%%%%%%%%%
%\addtolength{\textheight}{-9.35cm}
%%%%%%%%%%%%%%%%%%%%%%%%%%%%%%%%%%%%%%%%%%%%%%%%%%%%%%%%%%%%%%%%%%%%%
%% with an appropriate value. This command must be place on the second
%% last page, i.e., for a one-page abstract here, for a two-page
%% abstract right after the \maketitle command.

%% Create the title:
\maketitle

%% Abstract: 
%% For the final version of the accepted paper, please make sure you
%% remove the comment "THIS PAPER IS ELIGIBLE FOR THE STUDENT PAPER
%% AWARD."
%%
\begin{abstract}
	The paradigmatic view where information is seen as a more fundamental concept than the laws of physics leads to a different understanding of spacetime,
	where the causal order of events emerges from correlations between random variables representing physical quantities.
	In particular, such an information-theoretic approach does not enforce a global spacetime structure.
	By following this path, we conclude that perfect signaling correlations among three parties are possible which 
	do not obey the restrictions imposed 
	by global spacetime.
	We show this using
	a recent framework based on the sole assumptions that locally, quantum theory is valid and random variables can be described by probability distributions.
	Our result is of \emph{zero-error} type and can be seen as an analog to a tripartite appearance of quantum non-locality which manifests itself by satisfying a condition \emph{with certainty}, whereas the same is impossible for any local theory.
\end{abstract}

\section{Introduction}
Violations of Bell's inequality~\cite{Bell:1964ws} together with the existence of free randomness refute local realism---measurable quantities of quantum systems do not have locally predefined values.
To approach the problem of time, and to ultimately merge quantum theory and general relativity, this nonexistence of predefined values was applied to the causal order~of spacetime events~\cite{Hardy:2007bk}.
In the recent framework of Oreshkov, Costa, and Brukner~\cite{Oreshkov:2012uh}, no predefined causal order is assumed between experiments performed by different parties.
This relaxed setup allows for more general correlations.
In particular, they derive a causal inequality for two parties, violated by correlations that could not have been achieved using a predefined causal order.
Interestingly, this inequality has the same bound as Bell's inequality and is violated up to the same value as the maximal quantum-mechanical violation of Bell's inequality~\cite{Cirelson:1980}.
It has been partially proven that the achieved violation of the causal inequality~\cite{Oreshkov:2012uh} in indeed maximal~\cite{Brukner}.
%Note that the violation of the causal inequality is not claimed to be maximal~\cite{Oreshkov:2012uh}, however, we are aware of a partial proof by Brukner showing that the achieved violation is indeed maximal~\cite{Brukner}.
This similarity suggests a strong relation between non-locality, i.e., violations of Bell's inequality, and indefinite causal order.

This relation motivates us to study correlations with no predefined causal order between three parties.
As pointed out by Greenberger, Horne, and Zeilinger~\cite{Greenberger:1989vx,Greenberger:1990it}, deterministic non-local correlations of binary variables are possible for three parties or more.
Here, we tighten this relationship between non-locality and indefinite causal order by showing its analog:
Deterministic signaling correlations with indefinite causal order among three parties are possible.

This result is elaborated after a discussion of causal order, followed by a description of the framework for indefinite causal order~\cite{Oreshkov:2012uh}.
We show the result in two steps.
First, we present tripartite games and calculate the winning probability under the assumption of definite causal order.
In a next step, we show how the games are won with certainty using the mentioned framework.

\section{Definite and indefinite causal order}
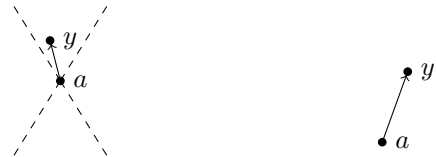
\begin{figure}
	\begin{center}
		\subfloat[Signaling from~$a$ to~$y$ depends on properties of a spacetime structure (here, signaling only into the future light-cone).\label{fig:signalingspacetime}]{
			\trimbox{-1.3cm 0cm -1.3cm 0cm}{
		\begin{tikzpicture}
			\node[draw,shape=circle,fill=black,minimum size=1mm,inner sep=0pt,outer sep=0pt] (C) {};
			\node[right=0mm of C] (a) {$a$};
			\draw[-,dashed] (C.center)--++(0.618,1);
			\draw[-,dashed] (C.center)--++(-0.618,1);
			\draw[-,dashed] (C.center)--++(0.618,-1);
			\draw[-,dashed] (C.center)--++(-0.618,-1);
			\node[draw,shape=circle,fill=black,minimum size=1mm,inner sep=0pt,outer sep=0pt,above left=5mm and 1mm of C.center] (Y) {};
			\node[right=0mm of Y] (y) {$y$};
			\draw[->] (C) -- (Y);
		\end{tikzpicture}}}
		\qquad
		\subfloat[Signaling from~$a$ to~$y$ depends on the intrinsic property that~$a$ is free (no assumption of spacetime).\label{fig:signalinginfo}]{
			\trimbox{-1.3cm 0cm -1.3cm 0cm}{
		\begin{tikzpicture}
			\node[draw,shape=circle,fill=black,minimum size=1mm,inner sep=0pt,outer sep=0pt] (C) {};
			\node[right=0mm of C] (a) {$a$};
			\node[draw,shape=circle,fill=black,minimum size=1mm,inner sep=0pt,outer sep=0pt,above right=9mm and 3mm of C.center] (Y) {};
			\node[right=0mm of Y] (y) {$y$};
			\draw[->] (C) -- (Y);
		\end{tikzpicture}}}
		\caption{Physically motivated (a) and information-based (b) approach to define the signaling direction.}
		\label{fig:signaling}
	\end{center}
\end{figure}

Physical quantities are described by random variables equipped with spacetime locations.
The locations specify when and where the variables are drawn.
Signaling from one location to another is then expressed by correlations among the respective variables.
However, correlations are merely a necessary, but not sufficient, condition for signaling.
To obtain signaling correlations, one of the correlated variables needs to be chosen freely.
This introduces the missing asymmetry and specifies the direction of signaling.
Usually, an underlying spacetime structure, e.g., relativistic spacetime, is assumed, and based on that, free choice is defined~\cite{Colbeck:2011hw,Ghirardi:2013bb} (see Figure~\ref{fig:signalingspacetime}).

Here, the often implicitly made assumption of a global spacetime is dropped.
Spacetime rather emerges from the observed correlations.
So, we cannot use its features to define free choice.
Instead, free choice is an intrinsic property of the random variables (see Figure~\ref{fig:signalinginfo}).
This alternative approach follows the trend of placing information theory as fundament for physical theories~\cite{Caves:2002hf,Brukner:2003,Clifton:2003di,Brassard:2005di,Pawiowski:2009dt,Chiribella:2011jb,Muller:2013gn,Pfister:2013ik}.

A party~$A$ has a local time~$t^A$ and is equipped with a set~$\Gamma^A=\{(X_i,\Omega_i)\}_I$
of random variables~$X_i$ with respective sample spaces~$\Omega_i$.
Note that for all~$i\in I$, the probability~$P(\Omega_i)$ is unity.
We use the event~$\Omega_i$ to define the local time~$t^A_i$, that is, the random variable~$X_i$ takes a fixed value at time~$t^A=t^A_i$.
Thus, the set~$\Gamma^A$ is totally ordered.
The properties of a totally ordered set are
1) antisymmetry, i.e., for all~$i$ and $j$, if~$X_i\preceq X_j$ and~$X_i\succeq X_j$, then~$X_i=X_j$,
2)~transitivity, i.e., for all~$i$,~$j$ and~$k$, if~$X_i\preceq X_j$ and~$X_j\preceq X_k$, then~$X_i\preceq X_k$, and
3) totality, i.e., for all~$i$,~$j$, either~$X_i\preceq X_j$ or~$X_i\succeq X_j$.
Note that totality implies reflexivity, i.e., for all~$i$,~$X_i\preceq X_i$.
We say a random variable~$X_i$ is in the \emph{past} of a random variable~$X_j$, denoted by~$X_i\preceq X_j$, if and only if~$t^A_i \le t^A_j$.
Alternatively,~$X_i$ is in the \emph{future} of~$X_j$, denoted by~$X_i\succeq X_j$, if and only if~$t^A_i \ge t^A_j$.
Assume for simplicity that no two random variables take a fixed value at the same local time, i.e., for all~$i\not= j$, we have~$t^A_i\not=t^A_j$.

Let us introduce a second party~$B$ with a local time~$t^B$ and a totally ordered set~$\Gamma^B=\{(Y_j,\chi_j)\}_J$ of random variables~$Y_j$ with respective sample spaces~$\chi_j$.

We do not assume a global time.
Thus, the local time~$t^A$ of~$A$ cannot be compared to the local time~$t^B$ of~$B$.
Nevertheless, we can causally order the random variables of~$A$ and of~$B$.
For that purpose, we use the notion of \emph{free choice}: If a random variable~$X_i$ is \emph{free}, then it can only be correlated with random variables in its causal future.
We say a random variable~$X_i$ is in the \emph{causal past} of another random variable~$Y_j$, if and only if~$X_i$ and~$Y_j$ are correlated and $X_i$ is free.
As above, we denote this by~$X_i\preceq Y_j$.
Conversely,~$X_i$ is in the \emph{causal future} of~$Y_j$, denoted by~$X_i\succeq Y_j$, if and only if~$X_i$ and~$Y_j$ are correlated and~$Y_j$ is free.
If~$X_i$ and~$Y_j$ are either not correlated or they are correlated but none of them is free, then~$X_i$ and~$Y_j$ are said to be \emph{separated}, denoted by~$X_i\not\prec\not\succeq Y_j$.
Due to this last property, a causal order does not satisfy totality, but remains reflexive, and thus is a partial order.

Currently, we are aware of two definitions for \emph{definite causal order}.
We call a causal order \emph{convex-definite}, if and only if it can be expressed by a convex combination of partial orders.
Another definition~\cite{FabioChristina}, which we call \emph{adaptive-definite}, is a convex combination of partial orders, whereas each party can arbitrarily choose the causal order between the parties in its causal future.
If a causal order is neither convex-definite nor adaptive-definite, then we call it \emph{indefinite}.

\section{Framework for local quantum mechanics with indefinite causal order}
The framework for local quantum mechanics with indefinite causal order by Oreshkov, Costa, and Brukner~\cite{Oreshkov:2012uh} models correlations between parties that do not share a global time.
It unifies no-signaling and signaling correlations.
In quantum mechanics, no-signaling correlations arise by local measurements on a shared quantum state, and signaling correlations arise by encoding information into a quantum system that subsequently is sent to another party via a quantum channel.

This framework models a party as a closed laboratory with a local time.
Each laboratory is opened once allowing a quantum system to pass.
During the passage, the quantum system undergoes a quantum-mechanical evolution chosen by the party.

The assumptions of the framework are
1) free choice, i.e., variables can be intrinsically free,
2) closed laboratories, i.e., a party can only receive a bit during the single opening of the laboratory,
and, 
3) local quantum mechanics, i.e., quantum mechanics is valid inside the laboratories.

Consider the bipartite case with parties~$A$ and~$B$.
Let the variables of~$A$ be~$\{a,x\}$, where~$a$ is free.
The variables of~$B$ are~$\{b,y\}$, with~$b$ being free.
Denote the input, respectively output, Hilbert space of~$A$ by~$A_1$,~$A_2$.
Analogously, the input/output Hilbert spaces of~$B$ are~$B_1$,~$B_2$.
Using the Choi-Jamio{\l}kowsky (CJ) representation, we can express the quantum-mechanical evolutions as objects on~$A_1\otimes A_2$ for~$A$, and on~$B_1\otimes B_2$ for~$B$. 
The most general quantum-mechanical evolutions are described by completely positive (CP) trace-nonincreasing maps.
In particular, CP maps can produce a classical outcome.
Let~$x$ and~$y$ be these outcomes for~$A$ and~$B$ respectively.
Then,~$A$'s map depending on the free choice~$a$ and yielding outcome~$x$ is~$M^{A_1,A_2}_{x,a}$.
The CP map of~$B$ is~$M^{B_1,B_2}_{y,b}$.
The maps~$\sum_x M^{A_1,A_2}_{x,a}$ and~$\sum_y M^{B_1,B_2}_{y,b}$ are completely positive trace-preserving (CPTP), because no classical outcome is produced.

\begin{figure}
	\begin{center}
		\begin{tikzpicture}
			\node[draw,rectangle,minimum width=0.5cm,minimum height=1cm] (A) {$A$};
			\node[draw,rectangle,minimum width=1cm,minimum height=1cm,right=of A] (W) {$W$};
			\node[draw,rectangle,minimum width=0.5cm,minimum height=1cm,right=of W] (B) {$B$};
			\draw[->] (A) to[out=80,in=100] (W);
			\draw[->] (B) to[out=100,in=80] (W);
			\draw[<-] (A) to[out=280,in=260] (W);
			\draw[<-] (B) to[out=260,in=280] (W);
			\draw[->] (A.150) -- ++(-0.5,0) node [left] {$x$};
			\draw[<-] (A.210) -- ++(-0.5,0) node [left] {$a$};
			\draw[->] (B.30) -- ++(0.5,0) node [right] {$y$};
			\draw[<-] (B.330) -- ++(0.5,0) node [right] {$b$};
		\end{tikzpicture}
		\caption{Process matrix as backward in time channel.}
		\label{fig:process}
	\end{center}
\end{figure}
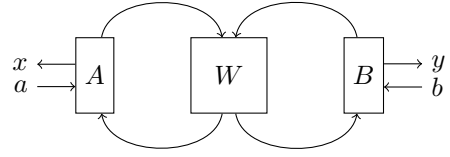

The probability of observing~$x$ and~$y$, given the free variables, is a bilinear function of the corresponding CP maps.
Thus, it can be expressed as
\begin{align}
\Pr(x,y|a,b)=\Tr\left[\left(M^{A_1,A_2}_{x,a} \otimes M^{B_1,B_2}_{y,b}\right)
W^{A_1,A_2,B_1,B_2}
\right]
\,,
\end{align}
where the so called process matrix~$W^{A_1,A_2,B_1,B_2}$ is an object in~$A_1\otimes A_2\otimes B_1\otimes B_2$.
The process matrix can be thought of as a backward in time channel (see Figure~\ref{fig:process}).
Requiring for each choice of the CP maps that the probability~$\Pr(x,y|a,b)$ is non-negative and sums up to unity, i.e., is unity for any choice of CPTP maps, gives three restrictions on the space of~$W^{A_1,A_2,B_1,B_2}$.
1) Valid process matrices can be written in the form~$W^{A_1,A_2,B_1,B_2}=a_0\id^{A_1,A_2,B_1,B_2}+\sum_{i>0}a_iW^{A_1,A_2,B_1,B_2}_i$, where, for each~$i$,~$a_i$ is a number and the matrix~$W^{A_1,A_2,B_1,B_2}_i$ is traceless and has at least one party with identity on the output Hilbert space and something different from identity on the input Hilbert space.
In particular, this excludes causal loops.
Furthermore, 2) process matrices are positive semi-definite, and 3) have trace~$d_{A_2}d_{B_2}$, where~$d_{A_2}$,~$d_{B_2}$ is the dimension of~$A_2$,~$B_2$.
This structure of valid process matrices is preserved when extended to more than two parties.

\section{Tripartite causal inequalities}
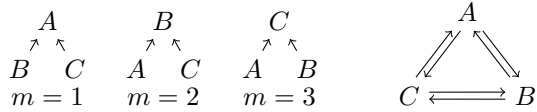
\begin{figure}
	\begin{center}
	\subfloat[][All-to-one signaling game. Depending on~$m$, party~$A$,~$B$, or~$C$ gets the parity the other parties' input.\label{fig:alltoone}]{
	\begin{tikzpicture}
		\node (A1) {$A$};
		\node[below of=A1] {$m=1$};
		\node[below left=0.4cm and 0.1cm of A1.center] (B1) {$B$};
		\node[below right=0.4cm and 0.1cm of A1.center] (C1) {$C$};
		\draw[->] (B1) -- (A1);
		\draw[->] (C1) -- (A1);
		\node[right=1cm of A1] (B2) {$B$};
		\node[below of=B2] {$m=2$};
		\node[below left=0.4cm and 0.1cm of B2.center] (A2) {$A$};
		\node[below right=0.4cm and 0.1cm of B2.center] (C2) {$C$};
		\draw[->] (A2) -- (B2);
		\draw[->] (C2) -- (B2);
		\node[right=1cm of B2] (C3) {$C$};
		\node[below of=C3] {$m=3$};
		\node[below left=0.4cm and 0.1cm of C3.center] (A3) {$A$};
		\node[below right=0.4cm and 0.1cm of C3.center] (B3) {$B$};
		\draw[->] (A3) -- (C3);
		\draw[->] (B3) -- (C3);
	\end{tikzpicture}}
	\qquad
	\subfloat[][Selective signaling game. Depending on $m$, one party has to signal to another.\label{fig:selectivesignaling}]{
	\begin{tikzpicture}
		\node (A) {$A$};
		\node[below right=0.86cm and 0.5cm of A.center] (B) {$B$};
		\node[below left=0.86cm and 0.5cm of A.center] (C) {$C$};
		\draw[->] (A.310) -- (B.110);
		\draw[<-] (A.290) -- (B.130);
		\draw[->] (B.190) -- (C.350);
		\draw[<-] (B.170) -- (C.10);
		\draw[->] (C.70) -- (A.230);
		\draw[<-] (C.50) -- (A.250);
	\end{tikzpicture} }
	\caption{Tripartite games.}
	\label{fig:tripartitegames}
	\end{center}
\end{figure}
We present two causal inequalities for tripartite games.
Both games are depicted in Figure~\ref{fig:tripartitegames}.
The inequalities hold under the assumptions of 1) free choice, 2) closed laboratories, and 3) definite causal order.
The previously described framework violates these inequalities up to their algebraic maximum.
Because each laboratory is opened only once, the sending and receiving process happens during this single opening.
This allows us to partially order the parties.
Consider the parties~$K$ and~$L$ who open their laboratories once.
Let~$\{k_i\}_I$, respectively~$\{\ell_j\}_J$, be the random variables of~$K$,~$L$.
If there exists a~$i\in I$ and a~$j \in J$, such that~$k_i\preceq \ell_j$, then we write~$K\preceq L$.
Alternatively, if there exists a~$i'\in I$ and a~$j'\in J$, such that~$k_{i'}\succeq \ell_{j'}$, then we write~$K\succeq L$.
There exists no pairs~$i,j$ and~$i',j'$, such that~$k_i\preceq \ell_j$ and~$k_{i'}\succeq \ell_{j'}$, as this would require a laboratory to open more than once.

Let the parties~$A$,~$B$, and~$C$ have the respective random variables~$\{a,x,m\}$,~$\{b,y,m\}$, and~$\{c,z,m\}$, where~$a$,~$b$,~$c$, and~$m$ are free and uniformly distributed.
All random variables except~$m$ are bits.
The shared variable~$m$ can take three values in the first game, and six values in the second game.

\subsection{All-to-one signaling}
In this game (see Figure~\ref{fig:alltoone}), a random party is selected to receive the parity of the other parties' input.
The winning probability, subject to maximization, is
\begin{align}
p_\text{succ}:=\frac{1}{3}
&\left(
\Pr(x=b\oplus c|m=1)
+
\Pr(y=a\oplus c|m=2)
\right.\notag\\
&\left.
+
\Pr(z=a\oplus b|m=3)
\right)
\,,
\end{align}
where the symbol~$\oplus$ denotes sum modulo~2.
Under the described assumptions, the winning probability cannot reach unity.

We calculate the bound under the assumption of a convex-definite causal order.
Unless~$B\preceq A$ and~$C\preceq A$, the probability~$\Pr(x=b\oplus c|m=1)$ is one-half, i.e., party~$A$ can only randomly guess the parity.
For the probability conditioned by~$m=2$ to be different from one-half, we require~$A\preceq B$ and~$C\preceq B$.
In the third case,~$m=3$, the parties have to be ordered by~$A\preceq C$ and~$B\preceq C$ to get a different value from one-half.
The requirements on the causal order in each case,~$m=1,2,3$, mutually contradict.
Thus, by fulfilling one requirement, the other two probability expressions are forced to be one-half.
This gives the upper bound
\begin{align}
p_\text{succ} \le \frac{1}{3}\left(1+\frac{1}{2}+\frac{1}{2}\right)=\frac{2}{3}
\,.
\end{align}
Using the adaptive definition for definite causal order, we get the following bound.
Without loss of generalization assume~$A$ is in the causal past of~$B$ and~$C$, i.e.,~$A\preceq B$ and~$A\preceq C$.
Then,~$A$ can arbitrarily choose between the orderings~$B\preceq C$ and~$B\succeq C$.
If~$m=2$,~$A$ chooses~$B\succeq C$.
Together with the assumption~$A\preceq C$, the requirements are fulfilled for the expression~$\Pr(y=a\oplus c|m=2)$ to reach unity.
If~$m=3$,~$A$ chooses the order~$B\preceq C$, allowing~$\Pr(z=a\oplus b|m=3)$ to reach unity.
Only in the last case~$m=1$, the value of the probability~$\Pr(x=b\oplus c|m=1)$ is one-half, as we assumed~$A$ is first.
This gives the bound
\begin{equation}
	p_\text{succ} \le \frac{1}{3}\left( 1+1+\frac{1}{2} \right)=\frac{5}{6}
	\,.
\end{equation}

\subsection{Selective signaling between any two parties}
In the second game (see Figure~\ref{fig:selectivesignaling}), a random sender and a random receiver are selected.
The game is won if the sender can perfectly send its bit to the receiver.
The game winning probability is
\begin{align}
q_\text{succ}:=
\frac{1}{6}
&\left(\Pr(y=a|m=1)+\Pr(z=a|m=2)\right.\notag\\
&\left.+\Pr(x=b|m=3)+\Pr(z=b|m=4)\right.\notag\\
&\left.+\Pr(x=c|m=5)+\Pr(y=c|m=6)\right)
\,.
\end{align}

We start by showing the upper bound in a scenario with a convex-definite causal order.
The case~$m=1$ needs~$A\preceq B$ for the probability to be different from one-half, whereas~\mbox{$m=3$} requires~$A\succeq B$, leading to a contradiction.
Furthermore, the case~$m=2$ contradicts with the case~$m=5$, and the case~$m=4$ contradicts with the case~$m=6$.
Thus, allowing one probability expression to be different from one-half forces another probability expression to be one-half.
This gives the bound
\begin{align}
q_\text{succ}\le \frac{1}{6}\left(3+3\cdot\frac{1}{2}\right)=\frac{3}{4}
\,.
\end{align}
Using an adaptive-definite causal order, we again assume without loss of generality that~$A$ is first.
Then, the bound is
\begin{equation}
	q_\text{succ} \le \frac{1}{6}\left( 4+2\cdot\frac{1}{2} \right)=\frac{5}{6}
	\,,
\end{equation}
as all probability terms are unity, except when~$A$ has to guess another party's bit, i.e., for~$m=3$ and~$m=5$.

\section{Maximally violating the inequalities}
In the following, we present the process matrix and the strategies to maximally violate both inequalities.
\subsection{Process matrix}
Denote by~$A_1$,~$B_1$, and~$C_1$ the input Hilbert spaces of the parties~$A$,~$B$, and~$C$, respectively.
The output Hilbert spaces are~$A_2$,~$B_2$, and~$C_2$.
These Hilbert spaces are two-dimensional.
Let
\begin{align}
o_1&:=
\id^{A_1}\otimes
\sigma_z^{A_2}\otimes
\sigma_z^{B_1}\otimes
\id^{B_2}\otimes
\sigma_z^{C_1}\otimes
\sigma_z^{C_2}\\
o_2&:=
\sigma_z^{A_1}\otimes
\id^{A_2}\otimes
\sigma_x^{B_1}\otimes
\sigma_z^{B_2}\otimes
\sigma_y^{C_1}\otimes
\sigma_z^{C_2}\\
o_3&:=
\sigma_z^{A_1}\otimes
\sigma_z^{A_2}\otimes
\sigma_y^{B_1}\otimes
\sigma_z^{B_2}\otimes
\sigma_x^{C_1}\otimes
\id^{C_2}
\,,
\end{align}
where~$\sigma_x$,~$\sigma_y$, and~$\sigma_z$ are the Pauli matrices.
This process matrix
\begin{align}
W^{A_1,A_2,B_1,B_2,C_1,C_2}=\frac{1}{8}
\left(
\id
+o_1+o_2+o_3
\right)
\end{align}
can be used to win the both games perfectly.
In the following, we will use shorthand~$W$ to denote~$W^{A_1,A_2,B_1,B_2,C_1,C_2}$.

First, let us verify that this process matrix is valid.
In~$o_1$,~$B$'s part is~$\sigma_z^{B_1}\otimes\id^{B_2}$.
In~$o_2$ and~$o_3$, we respectively have~$\sigma_z^{A_1}\otimes\id^{A_2}$ and~$\sigma_x^{C_1}\otimes\id^{C_2}$,
fulfilling the first requirement for~$W$ to be valid.
The terms~$o_1$,~$o_2$, and~$o_3$ are traceless.
Therefore,~$\Tr\left[W\right]=2^3$, fulfilling the third requirement.
Finally, we show that~$W$ is positive semi-definite.
All three terms~$o_1$,~$o_2$, and~$o_3$ mutually commute.
This implies a common set of eigenvectors~$\{\ket v\}_v$, i.e.,
\begin{align}
o_1\ket v&=\lambda_1^v\ket v\\
o_2\ket v&=\lambda_2^v\ket v\\
o_3\ket v&=\lambda_3^v\ket v
\,,
\end{align}
where~$\lambda_i^v$ are the corresponding eigenvalues.
Note the relation~$o_1\cdot o_2=o_3$.
Multiplying the first equation by~$o_2$ gives
\begin{align}
o_2o_1\ket v=\lambda_1^v o_2\ket v
\end{align}
which is
\begin{align}
o_3\ket v=\lambda_1^v\lambda_2^v\ket v
\,.
\end{align}
From this follows that for a given eigenvector~$\ket v$, the eigenvalues of~$o_3$ are determined by the eigenvalues of~$o_1$ and~$o_2$
\begin{align}
\lambda_3^v=\lambda_1^v\lambda_2^v
\,.
\end{align}
Because the terms~$o_1$ and~$o_2$ contain only Pauli matrices, their eigenvalues are~$1$ and~$-1$.
Therefore, the minimal eigenvalue of~$\id+o_1+o_2+o_3$ is zero.

\subsection{Local strategies}
The CP maps used to win the games, describe a measurement of the system on the input Hilbert space followed by a construction of the system on the output Hilbert space.
Using the CJ representation, each map used in the strategies can be written as
\begin{align}
R^{H,\alpha}_{i,k}:=
\left(\frac{\id+(-1)^i\sigma_\alpha}{2}\right)^{H_1}
\otimes
\left(\frac{\id+(-1)^k\sigma_z}{2}\right)^{H_2}
\,,
\end{align}
where~$H_1$ and~$H_2$ describe the corresponding input/output Hilbert space, and~$\alpha\in\{x,y,z\}$ is the measurement direction.
The measurement outcome is assigned to~$i$.
The outgoing system encodes~$k$ in~$z$ direction.

\subsubsection{Perfectly win all-to-one signaling game}
The CP maps in the CJ representation to win the all-to-one signaling game are presented in Table~\ref{tab:strategiesall-to-one}.
\begin{table}
\begin{center}
\begin{tabular}{l|lll}
$m$&$A$&$B$&$C$\\
\hline
$m=1$	&	$R^{A,z}_{x,a}$		&	$R^{B,x}_{y,b+y}$	&	$R^{C,y}_{z,c+z}$\\[1ex]
$m=2$	&	$R^{A,z}_{x,a}$		&	$R^{B,z}_{y,b}$			&	$R^{C,z}_{z,c+z}$\\[1ex]
$m=3$	&	$R^{A,z}_{x,a+x}$	&	$R^{B,y}_{y,b+y}$		&	$R^{C,x}_{z,c}$	
\end{tabular}
\end{center}
\caption{Strategies to win the all-to-one game.}
\label{tab:strategiesall-to-one}
\end{table}

For~$m=1$, the probability distribution of the joint outcomes, given the free variables, is
\begin{align}
\Pr&(x,y,z|a,b,c,m=1)\\
&=\Tr\left[\left(R^{A,z}_{x,a}\otimes R^{B,x}_{y,b+y}\otimes R^{C,y}_{z,c+z}\right)W\right]\\
&=\frac{1}{8}\left(1+(-1)^{x+b+c}\right)
\,.
\end{align}
Thus, the probability for~$A$ to receive~$b\oplus c$ is
\begin{align}
\Pr&(x=b\oplus c|a,b,c,m=1)\\
&=\sum_{y,z}\Pr(x=b\oplus c,y,z|a,b,c,m=1)\\
&=4\frac{1}{8}(1+1)=1
\,.
\end{align}
The probability distributions of the joint outcomes for~$m=2$, and~$m=3$ are
\begin{align}
\Pr&(x,y,z|a,b,c,m=2)\\
&=\Tr\left[\left(R^{A,z}_{x,a}\otimes R^{B,z}_{y,b}\otimes R^{C,z}_{z,c+z}\right)W\right]\\
&=\frac{1}{8}\left(1+(-1)^{a+y+c}\right)
\,,
\end{align}
and
\begin{align}
\Pr&(x,y,z|a,b,c,m=3)\\
&=\Tr\left[\left(R^{A,z}_{x,a+x}\otimes R^{B,y}_{y,b+y}\otimes R^{C,x}_{z,c}\right)W\right]\\
&=\frac{1}{8}\left(1+(-1)^{a+b+z}\right)
\,.
\end{align}
The probability for~$B$, respectively~$C$, to receive the parity of the other inputs is
\begin{align}
\Pr&(y=a\oplus c|a,b,c,m=2)\\
&=\sum_{x,z}\Pr(x,y=a\oplus c,z|a,b,c,m=2)\\
&=4\frac{1}{8}(1+1)=1
\,,
\\
\Pr&(z=a\oplus b|a,b,c,m=3)\\
&=\sum_{x,y}\Pr(x,y,z=a\oplus b|a,b,c,m=3)\\
&=4\frac{1}{8}(1+1)=1
\,.
\end{align}
Thus, the all-to-one game is won with certainty
\begin{align}
\frac{1}{3}(1+1+1)=1
\,.
\end{align}

\subsubsection{Perfectly win selective signaling between any two parties game}
To win the second game, the parties~$A$,~$B$, and~$C$ apply the maps according to Table~\ref{tab:strategiesselective}.
\begin{table}
\begin{center}
\begin{tabular}{l|lll}
$m$&$A$&$B$&$C$\\
\hline
$m=1$	&	$R^{A,z}_{x,a}$		&	$R^{B,z}_{y,b}$			&	$R^{C,z}_{z,z}$\\[1ex]
$m=2$	&	$R^{A,z}_{x,a+x}$	&	$R^{B,y}_{y,y}$			&	$R^{C,x}_{z,c}$\\[1ex]
$m=3$	&	$R^{A,z}_{x,a}$		&	$R^{B,x}_{y,y+b}$	&	$R^{C,y}_{z,z}$	\\[1ex]
$m=4$	&	$R^{A,z}_{x,x}$		&	$R^{B,y}_{y,y+b}$		&	$R^{C,x}_{z,c}$\\[1ex]
$m=5$	&	$R^{A,z}_{x,x}$		&	$R^{B,x}_{y,y}$			&	$R^{C,y}_{z,z+c}$\\[1ex]
$m=6$	&	$R^{A,z}_{x,0}$		&	$R^{B,z}_{y,b}$			&	$R^{C,z}_{z,z+c}$
\end{tabular}
\end{center}
\caption{Strategies to win the selective signaling game.}
\label{tab:strategiesselective}
\end{table}

We calculate the winning probabilities.
For~$m=1$, the probability that~$A$ can signal to~$B$ is~$\Pr(y=a|m=1)$.
This quantity is derived from the joint probability of the outcomes~$\Pr(x,y,z|a,b,c,m=1)$ which is
\begin{align}
\Tr\left[\left(
R^{A,z}_{x,a}\otimes
R^{B,z}_{y,b}\otimes
R^{C,z}_{z,z}
\right)W\right]
=\frac{1}{8}\left(1+(-1)^{a+y}\right)
\,.
\end{align}
The probability~$\Pr(y=a|m=1)$ thus is
\begin{align}
\sum_{x,z}\Pr(x,y=a,z|a,b,c,m=1)=4\frac{1}{8}(1+1)=1
\,.
\end{align}
The same holds for the other cases of~$m$.
Therefore, this game is won with certainty as well
\begin{align}
	\frac{1}{6}\left( 1+1+1+1+1+1 \right)=1
	\,.
\end{align}

\section{Conclusion and open questions}
We presented two games for three parties, that, under the assumption of definite causal order, cannot be won perfectly.
Both games ask for signaling correlations which cannot be fulfilled simultaneously.
Then, by dropping the assumption of definite causal order, we show that both games can be won with certainty.
For that purpose we use a recent framework for quantum correlations with no causal order~\cite{Oreshkov:2012uh}.
The correlations arising in the framework depend on local strategies and a resource called \emph{process matrix}.
We explicitly present the local strategies and the process matrix to perfectly win both games.
These deterministic correlations remind us of the result of Greenberger, Horne, and Zeilinger (GHZ)~\cite{Greenberger:1989vx,Greenberger:1990it}, where they achieve deterministic binary non-local correlations, whereas in the bipartite case, only non-deterministic binary non-local correlations are possible.
The GHZ result is insofar interesting, as the correlations are deterministic.
Variants of the GHZ result are Mermin's magic square~\cite{Mermin:1990he} and pseudo-telepathy~\cite{Brassard:1999,Brassard:2005}.
Furthermore, this idea of indefinite causal order was applied to quantum computation~\cite{Hardy:2009,Chiribella:2013bk}.

It would be very interesting to get a better understanding of the relationship between non-locality and indefinite causal order, and, in particular, to find a mapping from non-local games to non-causal games.
This would allow us to apply theorems from the more developed field of non-locality to the area of indefinite causal order.
Another question is whether such correlations appear in nature at all.

\section*{Acknowledgment}
We thank {\v C}aslav Brukner, Fabio Costa, Alberto Montina, and Jibran Rashid for helpful discussions.
This work was supported by the Swiss National Science Foundation and the National Centre of Competence in Research ``Quantum Science and Technology.''

%% References:
\bibliographystyle{IEEEtran}
\bibliography{refs}

% Generated by IEEEtran.bst, version: 1.13 (2008/09/30)
\begin{thebibliography}{10}
\providecommand{\url}[1]{#1}
\csname url@samestyle\endcsname
\providecommand{\newblock}{\relax}
\providecommand{\bibinfo}[2]{#2}
\providecommand{\BIBentrySTDinterwordspacing}{\spaceskip=0pt\relax}
\providecommand{\BIBentryALTinterwordstretchfactor}{4}
\providecommand{\BIBentryALTinterwordspacing}{\spaceskip=\fontdimen2\font plus
\BIBentryALTinterwordstretchfactor\fontdimen3\font minus
  \fontdimen4\font\relax}
\providecommand{\BIBforeignlanguage}[2]{{%
\expandafter\ifx\csname l@#1\endcsname\relax
\typeout{** WARNING: IEEEtran.bst: No hyphenation pattern has been}%
\typeout{** loaded for the language `#1'. Using the pattern for}%
\typeout{** the default language instead.}%
\else
\language=\csname l@#1\endcsname
\fi
#2}}
\providecommand{\BIBdecl}{\relax}
\BIBdecl

\bibitem{Bell:1964ws}
J.~S. Bell, ``{On the Einstein-Podolsky-Rosen paradox},'' \emph{Physics},
  vol.~1, no.~3, pp. 195--200, Nov. 1964.

\bibitem{Hardy:2007bk}
L.~Hardy, ``Towards quantum gravity: a framework for probabilistic theories
  with non-fixed causal structure,'' \emph{Journal of Physics A: Mathematical
  and Theoretical}, vol.~40, no.~12, pp. 3081--3099, Mar. 2007.

\bibitem{Oreshkov:2012uh}
O.~Oreshkov, F.~Costa, and {\v C}.~Brukner, ``{Quantum correlations with no
  causal order},'' \emph{Nature Communications}, vol.~3, Oct. 2012.

\bibitem{Cirelson:1980}
B.~Cirel'son, ``{Quantum generalizations of Bell's inequality},'' \emph{Letters
  in Mathematical Physics}, vol.~4, no.~2, pp. 93--100, Mar. 1980.

\bibitem{Brukner}
{\v C}.~Brukner, private communication, 2013.

\bibitem{Greenberger:1989vx}
D.~M. Greenberger, M.~A. Horne, and A.~Zeilinger, ``{Going beyond Bell's
  theorem},'' in \emph{Bell's Theorem, Quantum Theory, and Conceptions of the
  Universe}, M.~Kafatos, Ed.\hskip 1em plus 0.5em minus 0.4em\relax Dordrecht:
  Kluwer, 1989, pp. 69--72.

\bibitem{Greenberger:1990it}
D.~M. Greenberger, M.~A. Horne, A.~Shimony, and A.~Zeilinger, ``{Bell's theorem
  without inequalities},'' \emph{American Journal of Physics}, vol.~58, no.~12,
  pp. 1131--1143, 1990.

\bibitem{Colbeck:2011hw}
R.~Colbeck and R.~Renner, ``{No extension of quantum theory can have improved
  predictive power},'' \emph{Nature Communications}, vol.~2, p. 411, Aug. 2011.

\bibitem{Ghirardi:2013bb}
G.~Ghirardi and R.~Romano, ``{About possible extensions of quantum theory},''
  \emph{Foundations of Physics}, vol.~43, no.~7, pp. 881--894, Jun. 2013.

\bibitem{Caves:2002hf}
C.~Caves, C.~Fuchs, and R.~Schack, ``{Quantum probabilities as Bayesian
  probabilities},'' \emph{Physical Review A}, vol.~65, no.~2, p. 022305, Jan.
  2002.

\bibitem{Brukner:2003}
{\v C}.~Brukner and A.~Zeilinger, ``Information and fundamental elements of the
  structure of quantum theory,'' in \emph{Time, Quantum and Information},
  L.~Castell and O.~Ischebeck, Eds.\hskip 1em plus 0.5em minus 0.4em\relax
  Springer Berlin Heidelberg, 2003, pp. 323--354.

\bibitem{Clifton:2003di}
R.~Clifton, J.~Bub, and H.~Halvorson, ``{Characterizing quantum theory in terms
  of information-theoretic constraints},'' \emph{Foundations of Physics},
  vol.~33, no.~11, pp. 1561--1591, Nov. 2003.

\bibitem{Brassard:2005di}
G.~Brassard, ``{Is information the key?}'' \emph{Nature Physics}, vol.~1,
  no.~1, pp. 2--4, Oct. 2005.

\bibitem{Pawiowski:2009dt}
M.~Paw{\l}owski, T.~Paterek, D.~Kaszlikowski, V.~Scarani, A.~Winter, and
  M.~Zukowski, ``{Information causality as a physical principle},''
  \emph{Nature}, vol. 461, no. 7267, pp. 1101--1104, Oct. 2009.

\bibitem{Chiribella:2011jb}
G.~Chiribella, G.~M. D'Ariano, and P.~Perinotti, ``{Informational derivation of
  quantum theory},'' \emph{Physical Review A}, vol.~84, no.~1, p. 012311, Jul.
  2011.

\bibitem{Muller:2013gn}
M.~P. M{\"u}ller and L.~Masanes, ``Three-dimensionality of space and the
  quantum bit: an information-theoretic approach,'' \emph{New Journal of
  Physics}, vol.~15, no.~5, p. 053040, May 2013.

\bibitem{Pfister:2013ik}
C.~Pfister and S.~Wehner, ``An information-theoretic principle implies that any
  discrete physical theory is classical,'' \emph{Nature Communications},
  vol.~4, p. 1851, May 2013.

\bibitem{FabioChristina}
F.~Costa and C.~Giarmatzi, private communication, 2013.

\bibitem{Mermin:1990he}
N.~D. Mermin, ``{Quantum mysteries revisited},'' \emph{American Journal of
  Physics}, vol.~58, no.~8, p. 731, 1990.

\bibitem{Brassard:1999}
G.~Brassard, R.~Cleve, and A.~Tapp, ``Cost of exactly simulating quantum
  entanglement with classical communication,'' \emph{Physical Review Letters},
  vol.~83, pp. 1874--1877, Aug. 1999.

\bibitem{Brassard:2005}
G.~Brassard, A.~Broadbent, and A.~Tapp, ``{Recasting Mermin's multi-player game
  into the framework of pseudo-telepathy},'' \emph{Quantum Information \&
  Computation}, vol.~5, no.~7, pp. 538--550, Nov. 2005.

\bibitem{Hardy:2009}
L.~Hardy, ``Quantum gravity computers: on the theory of computation with
  indefinite causal structure,'' in \emph{Quantum Reality, Relativistic
  Causality, and Closing the Epistemic Circle}, ser. The Western Ontario Series
  in Philosophy of Science.\hskip 1em plus 0.5em minus 0.4em\relax Springer
  Netherlands, 2009, vol.~73, pp. 379--401.

\bibitem{Chiribella:2013bk}
G.~Chiribella, G.~M. D'Ariano, P.~Perinotti, and B.~Valiron, ``{Quantum
  computations without definite causal structure},'' \emph{Physical Review A},
  vol.~88, no.~2, p. 022318, Aug. 2013.

\end{thebibliography}

\end{document}